\documentclass[english,a4paper]{article}
\usepackage{dcolumn}
\usepackage{bm}
\usepackage{graphicx}
\usepackage{amsmath}
\usepackage{amsfonts}
\usepackage{amssymb}
\usepackage{amsthm}
\usepackage{amstext}
\usepackage{amsbsy}
\usepackage{amsopn}
\usepackage{amscd}
\usepackage{amsxtra}

\def\phi{\varphi}
\def\epsilon{\varepsilon}

\everymath{\displaystyle}

\begin{document}
\title{Time-optimal control of two-level quantum systems by piecewise constant pulses}
\author{E. Dionis\footnote{Laboratoire Interdisciplinaire Carnot de
Bourgogne (ICB), UMR 6303 CNRS-Universit\'e Bourgogne, 9 Av. A.
Savary, BP 47 870, F-21078 Dijon Cedex, France}, D. Sugny\footnote{Laboratoire Interdisciplinaire Carnot de
Bourgogne (ICB), UMR 6303 CNRS-Universit\'e Bourgogne, 9 Av. A.
Savary, BP 47 870, F-21078 Dijon Cedex, France, dominique.sugny@u-bourgogne.fr}}

\maketitle

\begin{abstract}
We apply an extension of the Pontryagin Maximum Principle to derive time-optimal controls of two-level quantum systems by means of piecewise constant pulses. Global optimal solutions are obtained for state-to-state transfer in the cases with one and two controls. Exact quantum speed limits are established as a function of the sampling period. We observe numerically an exponential convergence towards the minimum time in the continuous limit when this period goes to zero. We show that this convergence is only polynomial for a linearized quantum system. We discuss the experimental impact of this result.
\end{abstract}



\section{Introduction}
Designing fast and accurate control sequences while taking into account specific experimental constraints is a central task in quantum control~\cite{glaserreview,brifreview,roadmap,adiabaticreview,stefanatos2020,guery2019}. In this framework, optimal control theory (OCT)~\cite{pont,boscain21} is a powerful and versatile tool in order to propose efficient answers to the different issues that can be encountered in experimental setups~\cite{glaserreview,roadmap}. For complex quantum systems, optimal controls are computed by using iterative optimization algorithms~\cite{grape,gross,reichkrotov,calarco} such as the well-known GRAPE procedure~\cite{grape}.

A key limitation of pulse-shaping  devices is their ability to generate time-continuous controls. They are therefore usually replaced by piecewise constant pulses in experiments with a minimum sampling period over which the controls remain constant. This hybrid situation is characterized by a digital control which evolves in a discrete way, while the state of the system varies continuously. This constraint was one of the motivations for the development of GRAPE, a gradient-based algorithm directly optimizing piecewise constant functions that can be implemented experimentally. Even if the efficiency of numerical algorithms has been demonstrated in a variety of domains~\cite{glaserreview,roadmap}, their application is not completely satisfactory because only local optimal protocols are achieved, therefore with no certitude about the global optimality of the control process. This problem can be solved from the Pontryagin Maximum Principle (PMP), which has the decisive property of transforming the initial infinite-dimensional control problem into a generalized Hamiltonian system subject to a maximization condition and boundary constraints. The optimal solution corresponds then to a Hamiltonian trajectory both reaching the target state and minimizing a cost functional such as the control time. The low and finite dimension of this new dynamic space simplifies greatly the search for globally optimal controls~\cite{boscain21}. This method has been applied recently to a series of quantum control problems~\cite{boscain,alessandro,garon,hegerfeldt:2013,khaneja:2002,khanejaspin,lapert:2013,lapertprl}. However, as with the original formulation of Pontryagin~\cite{pont}, such studies consider that the control can be modified at any time, leading to continuous functions or possibly having discontinuities in the case of bang-bang protocols~\cite{boscain21}. As this assumption is too strong for standard wave-form generators, the PMP only gives mathematical optima which cannot necessarily be physically achieved in practice. An interesting example is given by time-optimal control processes in which the goal is to perform a given task as quickly as possible. In this framework, quantum speed limits provide lower bounds on the minimum time required for a given control protocol~\cite{QSLreview,QSLreview2,caneva2009}. Limiting the shaping to piecewise constant controls necessarily increases the minimum time. A key point for the experimental performance of the process is the quantitative impact of the control digitalization on this time.

In this study, we aim to take a step towards solving this control problem by applying a recent mathematical extension of the PMP to piecewise constant pulses~\cite{bourdin2016,bourdin2017,bourdin2021}. In this new and rigorous formulation (called \emph{discrete} below to differentiate it from the standard \emph{continuous} one), the overall structure of the PMP is not modified. The main difference between the continuous and the discrete versions relies on the maximization condition in which the instantaneous condition is replaced by an integral one. This integral condition can be interpreted as an average over the sampling period of the standard maximization condition. We apply this general procedure to the control of two-level quantum systems. We consider two benchmark control protocols with respectively one and two control parameters  and for which the time-optimal solutions are bang-bang and continuous. We find in the two cases the optimal dynamics for any sampling period and we show that the corresponding minimum time converges exponentially toward its continuous limit when this period goes to zero. We show that this convergence is only polynomial for a specific linearized quantum system. Finally, we describe the link between this new version of the PMP and GRAPE and we propose a new formulation of the algorithm based on the PMP with an exact gradient. A comparison with different versions of GRAPE is performed for a specific control problem. We consider for GRAPE both the split-operator~\cite{grape} and the auxiliary matrix~\cite{goodwin2015} approaches. In the first formulation, the gradient of the fidelity with respect to the control is approximated by replacing the elementary evolution operator over a time step into a product of two or more factors, while in the second the gradient is exactly computed by extending the size of the Hilbert space and computing the matrix exponential in this new space. We show the superiority of the PMP-based formulation for the example under study.

The paper is organized as follows. Section~\ref{sec2} introduces
the PMP for piecewise constant pulses. The time-optimal solutions for controlling two-level quantum systems with two and one control parameters are respectively presented in Sec.~\ref{sec3} and \ref{sec4}. The optimal solution for the control of a linear system is described in Sec.~\ref{seclinear}. Section~\ref{secgrape} focuses on the link between this new formulation of the PMP and gradient-based optimal algorithms. Conclusion and prospective views are
given in Sec.~\ref{sec5}. Technical computations are reported in
Appendix~\ref{appA}.

\section{Methodology}\label{sec2}
We briefly recall in this section the main steps of the application of the PMP for bilinear differential systems. We refer the interested reader to a recent paper about the PMP  in quantum control for details~\cite{boscain21}. We describe the differences in the PMP between the continuous and the discrete cases. We present a numerical method called the shooting algorithm to solve the optimization equations for low-dimensional dynamical systems. 

We consider the control of a system described by the state $X\in\mathbb{R}^n$, $n\geq 1$, whose dynamics are described by the following differential system
$$
\dot{X}=H_0X+u_1H_1X+u_2H_2X
$$
where $H_0$, $H_1$ and $H_2$ are $n\times n$ skew-symmetric real matrices, and $u_1(t)$, $u_2(t)$ the two real control parameters. Note that the approach detailed below can be straightforwardly extended to any number of control parameters. The goal of the control process is to bring in minimum time $t_f$ the state of the system from $X_0$ to the target state $X_f$ with the constraint $u_1(t)^2+u_2(t)^2\leq 1$ at any time $t\in [0,t_f]$.

We first state the PMP in the standard continuous situation where $u_1$ and $u_2$ are assumed to be sufficiently smooth functions~\cite{boscain21}. We only consider the regular case in which $u_1$ and $u_2$ are not simultaneously equal to 0, except in isolated points. The Pontryagin Hamiltonian $H_P$ can be expressed as
$$
H_P(X,P,u_1,u_2)=P^\intercal [H_0X+u_1H_1X+u_2H_2X]
$$
with the adjoint state $P\in\mathbb{R}^n$. Since the final time $t_f$ is free, $H_P$ is a constant fixed to 1 (see~\cite{boscain21} for technical details). The adjoint state $P$ fulfills the same equation as $X$
$$
\dot{P}=H_0P+u_1H_1P+u_2H_2P
$$
The candidate controls $u_1^*$ and $u_2^*$ for optimality satisfy the following maximization condition at any time $t$
$$
H(X,P,u_1^*(X,P),u_2^*(X,P))=\max_{u_1^2+u_2^2\leq 1}H_P(X,P,u_1,u_2)
$$
A straightforward computation shows that the solutions can be written as
\begin{eqnarray*}
& & u_1^*=\frac{P^\intercal H_1X}{\sqrt{(P^\intercal H_1X)^2+(P^\intercal H_2X)^2}}, \\
& & u_2^*=\frac{P^\intercal H_2X}{\sqrt{(P^\intercal H_1X)^2+(P^\intercal H_2X)^2}}.
\end{eqnarray*}
Hamiltonian trajectories of $H$ can then be computed from the expressions of $u_1^*$ and $u_2^*$. The shooting approach consists in finding the initial adjoint state $P(0)$ such that the corresponding trajectory reaches exactly the target state $X(t_f)=X_f$ in minimum time. The optimal controls are then deduced from the time evolution of $X$ and $P$. If the optimization equations cannot be integrated analytically, a Newton algorithm is generally used to estimate the initial adjoint state with a very high numerical precision.

We study now the case where $u_1$ and $u_2$ are piecewise constant pulses with the same constraint $u_1(t)^2+u_2(t)^2\leq 1$. We denote by $T$ the sampling period. The controls are characterized by $N$ values $u_1^{(k)}$, $u_2^{(k)}$ in the time intervals $[kT,(k+1)T]$, $k\in\{0,1,2,\cdots,(N-1)\}$, with $(N-1)T+\delta  T=t_f$ and $0<\delta T\leq T$. Note that the last time interval of length $\delta T$ can be lower or equal to $T$. This point is discussed in the different examples. We consider here for simplicity a regular time discretization but the same approach can be used if the sampling period is not constant.

The application of the PMP is very similar to the continuous version~\cite{bourdin2016,bourdin2017}. The Pontryagin Hamiltonian is defined the same way (except that $H_P$ is no more constant in time but still equal to +1 at the final time $t_f$) and the state $X$ and adjoint state $P$ satisfy the same differential equations. The only difference corresponds to the formulation of the maximization condition. Introducing the integrals
\begin{equation}\label{eqHk}
\begin{split}
&H_1^{(k)}=\int_{kT}^{(k+1)T}P(\tau)^\intercal H_1X(\tau)d\tau \\
&H_2^{(k)}=\int_{kT}^{(k+1)T}P(\tau)^\intercal H_2X(\tau)d\tau
\end{split}
\end{equation}
the maximization condition can be expressed for all $k$ as
$$
H_1^{(k)}(v_1-u_1^{(k)})+H_2^{(k)}(v_2-u_2^{(k)})\leq 0
$$
for any couple $(v_1,v_2)$ such that $v_1^2+v_2^2\leq 1$. We then deduce that the extremal controls are given by
\begin{equation}\label{equu}
\begin{split}
& u_1^{(k)*}=\frac{H_1^{(k)}}{\sqrt{(H_1^{(k)})^2+(H_2^{(k)})^2}}\\
& u_2^{(k)*}=\frac{H_2^{(k)}}{\sqrt{(H_1^{(k)})^2+(H_2^{(k)})^2}}.
\end{split}
\end{equation}
If the last time interval is not of length $T$ then the integrals~\eqref{eqHk} are taken from $(N-1)T$ to $t_f$. From this modified maximization condition, the same shooting approach can be used to find the initial value of $P$ bringing the system to the target state. Note that $\delta T$ is also computed by the shooting procedure when this parameter is not fixed. Finally, we point out that in the limit $T\to 0$, the piecewise constant controls converge towards their continuous limit given by the standard PMP~\cite{bourdin2016,bourdin2017}.
\section{The case of two controls}\label{sec3}
\subsection{The general formalism}
As a first application, we consider the time optimal control of a two-level quantum system by means of two resonant controls~\cite{boscain21}. In a given rotating frame, the dynamics of the system is given by the Schr\"odinger equation written in units such that $\hbar=1$
$$
i|\dot{\psi}\rangle=H|\psi\rangle=\frac{1}{2}(\omega_x\sigma_x+\omega_y\sigma_y)|\psi\rangle
$$
where $\omega_x$ and $\omega_y$ are the two real control amplitudes along the $x$ and $y$- directions with the constraint $\omega_x^2+\omega_y^2\leq 1$, and $\sigma_{x,y}$ the two Pauli matrices. We limit our study to state-to-state transfer. In the continuous case, it has been shown that the optimal protocol consists in saturating the bound at any time, that is $\omega_x^2+\omega_y^2=1$. This saturation can be qualitatively explained by the fact that a fastest control protocol requires generally maximum control intensity. The same assumption is made in this section for the discrete formulation of the PMP. We introduce to this aim the Bloch vector $X=(x,y,z)$ which satisfies
$$
\dot{X}=(\omega_xM_x+\omega_yM_y)X
$$
with the matrices $M_x$, $M_y$ and $M_z$ defined as
$$
M_x=\begin{pmatrix}
0 & 0 & 0\\
0 & 0 & -1\\
0 & 1 & 0
\end{pmatrix},~
M_y=\begin{pmatrix}
0 & 0 & 1\\
0 & 0 & 0\\
-1 & 0 & 0
\end{pmatrix},~
M_z=\begin{pmatrix}
0 & -1 & 0\\
1 & 0 & 0\\
0 & 0 & 0
\end{pmatrix}
$$
The Pontryagin Hamiltonian $H_P$ can be expressed as
$$
H_P=P^\intercal (\omega_xM_x+\omega_yM_y)X
$$
where the adjoint state $P=(p_x,p_y,p_z)$ fulfills the differential equation
$$
\dot{P}=(\omega_x M_x+\omega_yM_y)P.
$$
We consider piecewise constant pulses described by a phase $\varphi$ such that $\omega_x=\cos\varphi$ and $\omega_y=\sin\varphi$. We denote $\varphi_k$ the amplitude of the control parameter in the interval $[kT,(k+1)T]$, $0\leq k\leq (N-1)$. We also introduce $X_k$ and $P_k$ the values of the state and adjoint state at time $t=kT$, and the quantities $I_{x,y,z}^{(k)}=P_kM_{x,y,z}X_k$.
Using Eq.~\eqref{equu}, the maximization condition of $H_P$ leads to the condition
\begin{equation}\label{eqmax2}
\tan(\varphi_k)=\frac{\int_{kT}^{(k+1)T}P_k^\intercal(\tau)M_yX_k(\tau)d\tau}{\int_{kT}^{(k+1)T}P_k^\intercal(\tau)M_xX_k(\tau)d\tau}
\end{equation}
with $X_k(\tau)=R(\tau)X_k$ and $P_k(\tau)=R(\tau)P_k$, $\tau\in [kT,(k+1)T]$, and $R(\tau)=\exp[(\cos(\varphi_k)M_x+\sin(\varphi_k)M_y)\tau]$. The rotation matrix $R(\tau)$ can be explicitly computed as
$$
R(\tau)=\begin{pmatrix}
\sin^2(\varphi_k)\cos(\tau)+\cos^2(\varphi_k) & (1-\cos(\tau))\sin(\varphi_k)\cos(\varphi_k) & \sin(\tau)\sin(\varphi_k) \\
(1-\cos(\tau))\sin(\varphi_k)\cos(\varphi_k) & \sin^2(\varphi_k)+\cos(\tau)\cos^2(\varphi_k) & -\sin(\tau)\cos(\varphi_k) \\
-\sin(\tau)\sin(\varphi_k) & \sin(\tau)\cos(\varphi_k) & \cos(\tau)
\end{pmatrix}
$$
Setting $h_k=\tan(\varphi_k/2)$ and computing the integrals in Eq.~\eqref{eqmax2}, we arrive at a polynomial of order 2 in $h_k$
\begin{eqnarray*}
& & [I_y^{(k)}\sin(T)+(1-\cos(T)I_z^{(k)})]h_k^2+2I_x^{(k)}\sin(T)h_k\\
& & +[(1-\cos(T)I_z)-I_y\sin(T)]=0.
\end{eqnarray*}
The two real solutions (if they exist) are candidates to be the optimal value of $\varphi_k$. As in the continuous case, they lead to two symmetric trajectories with respect to the equator. We select in Sec.~\ref{numsec3} the positive root which corresponds to $p_z(0)<0$ and to a dynamic in the southern hemisphere. Finally, we point out that in the continuous limit $T\to 0$, we recover the extremal solution $\tan\varphi_k=\frac{I_y}{I_x}$.
\subsection{Numerical results}\label{numsec3}
We apply the shooting method to a state-to-state transfer on the Bloch sphere. The initial and target states are respectively of coordinates $(1,0,0)$ and $(0,1,0)$. As mentioned in Sec.~\ref{sec2}, we consider two cases for which $t_f$ is either equal to $NT$ or to $(N-1)T+\delta T$. In the first situation, for a fixed time step value, $N$, we only need to find the initial adjoint state $P(0)$ and the sampling period $T$ such that the trajectory reaches the target state in a minimum time $t_f$. In the numerical simulations, we therefore optimize four parameters, $p_x(0)$, $p_y(0)$, $p_z(0)$ and $T$ such that $x(t_f=NT)=0$, $y(t_f)=1$, $z(t_f)=0$ and $H_P(t_f)=1$. In the second case, $T$ is fixed and we find the coordinates of $P(0)$, $N$ and $t_f$ such that the four final conditions are met. The parameter $\delta T$ is then given by $\delta T=t_f-(N-1)T$. We point out that we use exactly $N$ time steps in the first approach, while the last step may be different from the others in the second case.

The continuous limit can be integrated exactly and leads to the minimum time $t_f^{(c)}=\frac{\pi\sqrt{3}}{2}\simeq 2.7207$ (see Ref.~\cite{boscain21} for the derivation). The corresponding initial adjoint state $P(0)$ has the coordinates $(p_x(0),\frac{1}{\sqrt{3}},\pm 1)$ where the first coordinate is not fixed. We stress that the same control process is obtained for any value of $p_x(0)$. Note that a similar behavior is observed in the discrete case. The two possible values of $p_z(0)$ correspond to two symmetric trajectories with respect to the equator on the Bloch sphere~\cite{boscain21}. Details about the results of the shooting method and a comparison with the continuous limit are given in Appendix~\ref{appA}.

Figure~\ref{fig1} compares the results obtained by the two different approaches. We observe a rapid convergence of the minimum time towards the continuous limit since the ratio between the two times is respectively of the order of $10^{-3}$ and $10^{-5}$ when $N=10$ and 100. The results of Fig.~\ref{fig1} clearly show that this convergence is exponential. As could be expected, releasing the constraint $t_f=NT$ allows to decrease the minimum time. In this second case, the curve $t_f$ as a function of $T$ has a quasi-quadratic shape when $T$ is not too small and horizontal tangents when $\delta T=T$.
\begin{figure}[tb]
  \centering
  \includegraphics[width=0.9\linewidth]{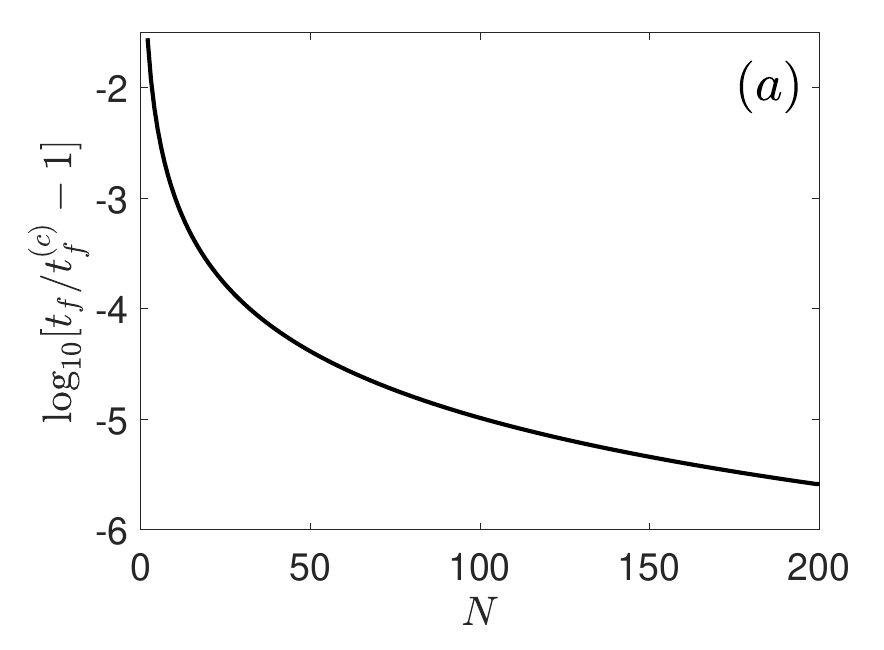}
  \includegraphics[width=0.9\linewidth]{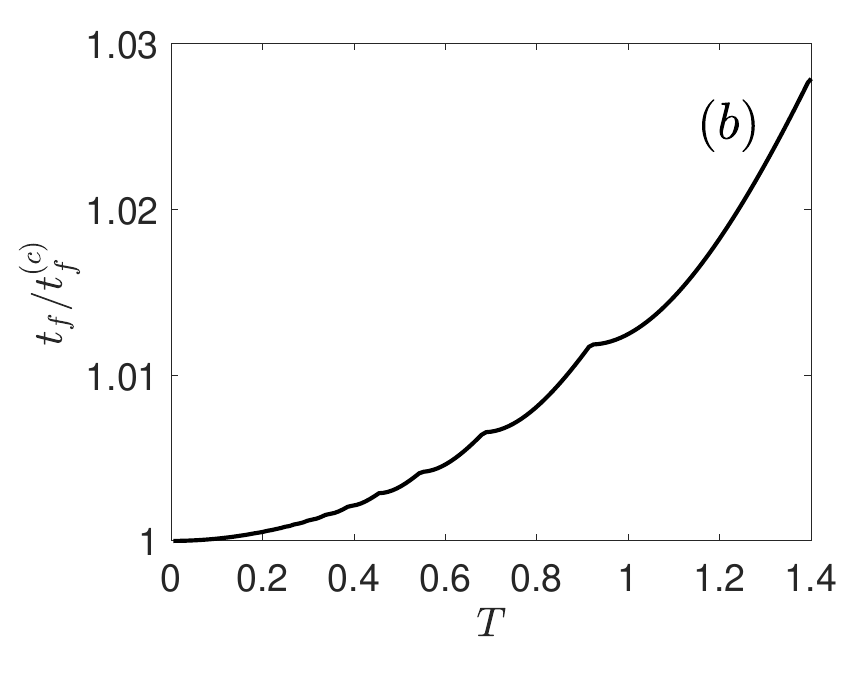}
   \caption{Evolution of the minimum time $t_f$ as a function of the number of time steps $N$ (a) (with $\delta T=T$) and of the sampling period $T$ (b) when $\delta T$ is free. In panel (a), the minimum value of $N$ is 2. The points are computed for integer values of $N$. The solid line is just to guide the reading. Quantities plotted are dimensionless.}
  \label{fig1}
\end{figure}
An example of optimal solutions is given in Fig.~\ref{fig2} for $N=3$ and $\delta T\neq T$. We observe that the discrete and the continuous trajectories are very close to each other on the Bloch sphere while the two controls are quite different even if they have a similar evolution.
\begin{figure}[tb]
  \centering
  \includegraphics[width=0.9\linewidth]{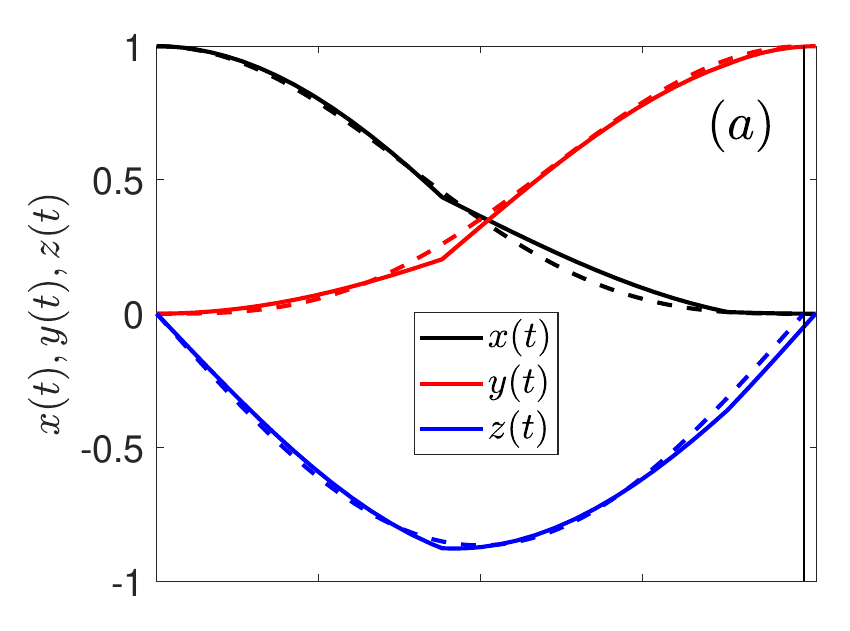}
  \includegraphics[width=0.9\linewidth]{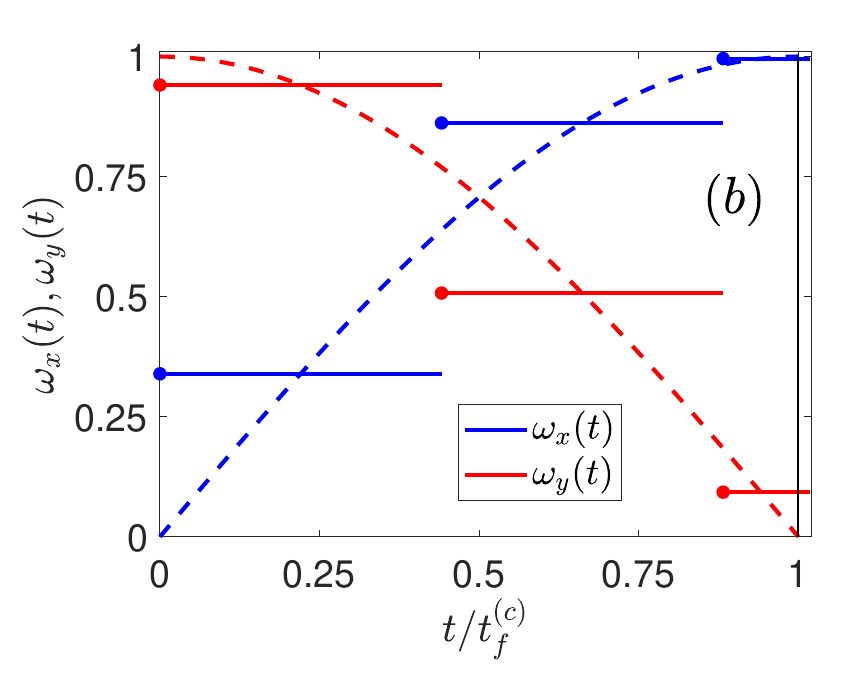}
   \caption{Panel (a) displays the optimal trajectories for $N=3$ and $\delta T\neq T$ (solid lines) and in the continuous limit (dashed lines) for the coordinates $x$ (black), $y$ (red or light gray) and $z$ (blue or solid gray). The black vertical line corresponds to the minimum continuous time $t_f^{(c)}$. The corresponding controls are represented in panel (b) with the same color code. The dots indicate the position of the switching times $t_k$. Quantities plotted are dimensionless.}
  \label{fig2}
\end{figure}

As a quantitative example, we briefly analyze the control of a spin 1/2 particle in Nuclear Magnetic Resonance~\cite{bernstein2004}. We consider the experimental parameters given in~\cite{kobzar2008}, i.e. a time digitization of 0.5~$\mu$s and a maximum amplitude of $\nu=100$~kHz. The experimental time $t_{\textrm{exp}}$ is then given by $t=2\pi\nu t_{\textrm{exp}}$ where $t$ is the time in normalized unit used in the theoretical part of Sec.~\ref{sec3}. For $N=9$ and $\delta T\neq T$, the minimum time is equal to 4.34~$\mu$s, while this time is 4.33~$\mu$s in the continuous limit. Due to the exponential convergence of the optimization procedure, we observe that the difference is very small between the two times even for relatively low values of $N$.

\section{The case of one control}\label{sec4}
\subsection{Description of the optimal control problem}\label{sec4a}
We consider in this section a similar control problem for a two-level quantum system, but with only one control $\omega(t)$ such that $|\omega(t)|\leq 1$~\cite{boscain,boscain21}. The dynamics of the Bloch vector are governed by the following differential equation
\begin{equation}\label{eqonecontrol}
\dot{X}=(\Delta M_z+\omega M_x)X
\end{equation}
where the offset $\Delta$ is a constant parameter and we introduce $\Omega=\sqrt{1+\Delta^2}$. The goal of the control is to steer the system from the north pole to the south one in minimum time. This corresponds to a complete population transfer from the ground state to the excited state in the two-level quantum system. The adjoint state $P$ fulfills the same equation as $X$ and the Pontryagin Hamiltonian $H_P$ is given by
$$
H_P=\Delta(p_yx-p_xy)+\omega(p_zy-p_yz)
$$
The continuous limit can be solved exactly with the PMP~\cite{boscain21,boscain}. In the case $|\Delta|\leq 1$, the optimal solution is a bang-bang control with a switching time equal to $t_1=\frac{1}{\Omega}(\pi-\arccos(\Delta^2))$ or $t_2=\frac{1}{\Omega}(\pi+\arccos(\Delta^2))$. These two symmetric solutions lead to the same total time $t_f=2\pi/\Omega$.
The discrete version of the PMP is applied in Appendix~\ref{appB}. The numerical results are described in Sec.~\ref{secnr}.

\subsection{Numerical results}\label{secnr}
\begin{figure}[tb]
  \centering
  \includegraphics[width=1\linewidth]{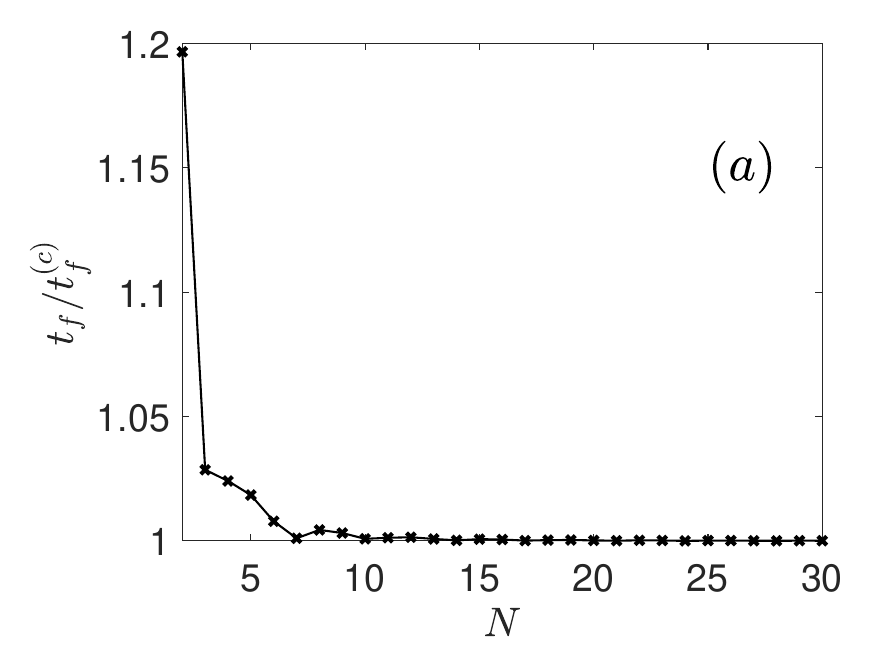}
  \includegraphics[width=1\linewidth]{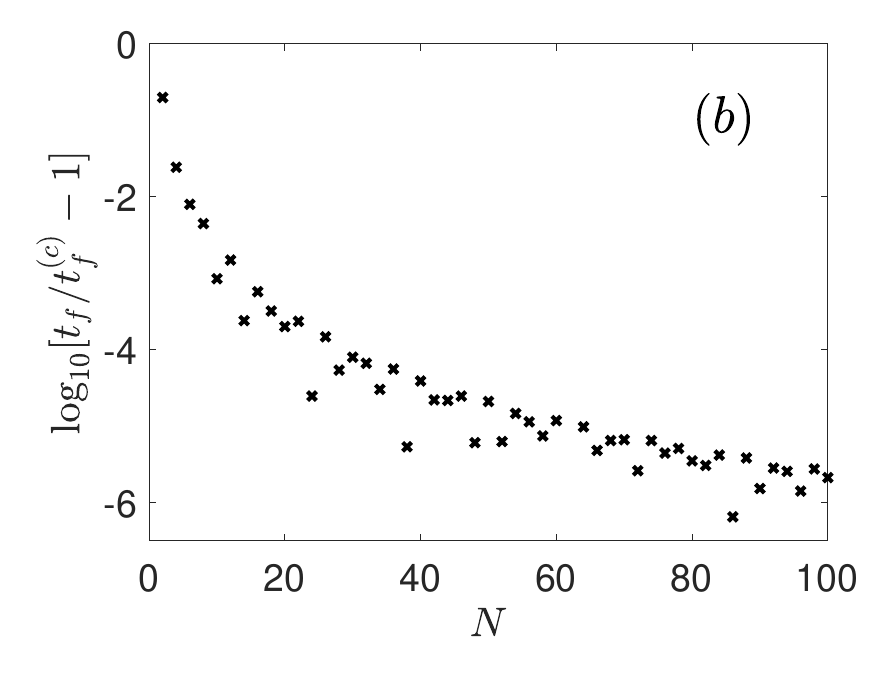}
   \caption{Evolution of the minimum time $t_f$ as a function of the number of steps $N$ (crosses). The solid line is just to guide the eye. The panel (b) corresponds to the same data as panel (a) but in a logarithmic scale. Quantities plotted are dimensionless.}
  \label{fig7}
\end{figure}

\begin{figure}[tb]
  \centering
  \includegraphics[width=1\linewidth]{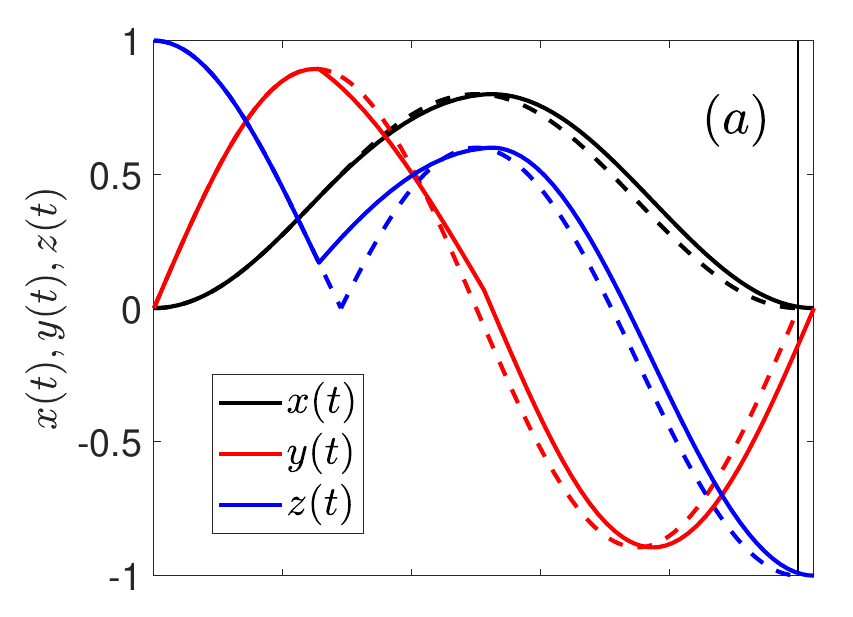}
  \includegraphics[width=1\linewidth]{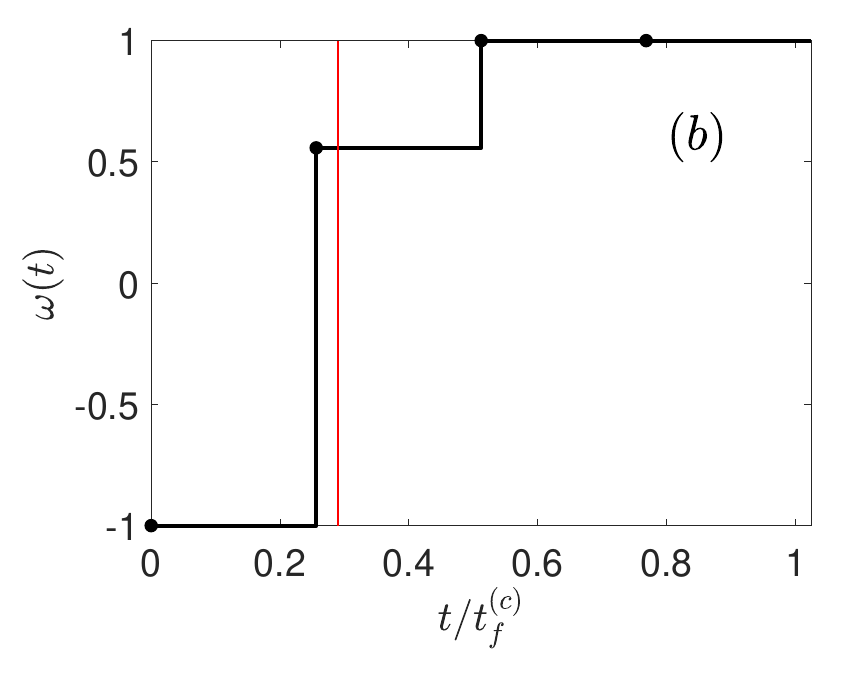}
   \caption{Panel (a): Optimal trajectories for $N=4$ and $\delta T=T$ (solid lines) and in the continuous limit (dashed lines) for the coordinates $x$ (black), $y$ (red or light gray) and $z$ (blue or solid gray). The black vertical line corresponds to the minimum continuous time $t_f^{(c)}$. Panel (b): Optimal control in the discrete case. The dots indicate the position of the times $t_k=kT$. The vertical red line represents the position of the switching time for the bang-bang solution in the continuous limit. Quantities plotted are dimensionless.}
  \label{fig4}
\end{figure}

The two shooting methods described in Sec.~\ref{sec2} can be applied to this control problem. We only present the results corresponding to $\delta T=T$. In this case, we observe numerically that the solution is very sensitive to the initial conditions of the adjoint vector. The initial guess was adapted to each value of $N$. Despite these numerical difficulties, we show in Fig.~\ref{fig7} that the minimum time $t_f$ converges very quickly towards its continuous limit as a function of $N$, the difference being of the order of $10^{-4}$ for $N=20$. The cloud of points that can be seen in Fig.~\ref{fig7}b is a signature of the numerical instabilities of the shooting algorithm, even if a clear trend can be observed with an approximate exponential convergence. An example of optimal trajectories and controls is plotted in Fig.~\ref{fig4}. The bang-bang optimal solution in the continuous limit is replaced by another piecewise constant control but with fixed switching times. While the former switches between -1 and 1 (singular control is not optimal in this case), the latter can take values in the interval $(-1,1)$ as allowed by the discrete version of the PMP. This degree of freedom compensates for the fact that switching cannot take place at any time $t\in [0,t_f]$. For most values of $N$, we observe that the algorithm finds the optimal control displayed in Fig.~\ref{fig4} with three constant parts of respective amplitudes -1, $\omega_0$ and +1. The control $\omega=\omega_0$ is applied during only one time step around the switching time in the continuous limit. This parameter $\omega_0\in [-1,1]$ depends in a rather complex way on $N$.

\subsection{Optimal control of a Landau-Zener-type Hamiltonian}
\begin{figure}[tb]
  \centering
  \includegraphics[width=1\linewidth]{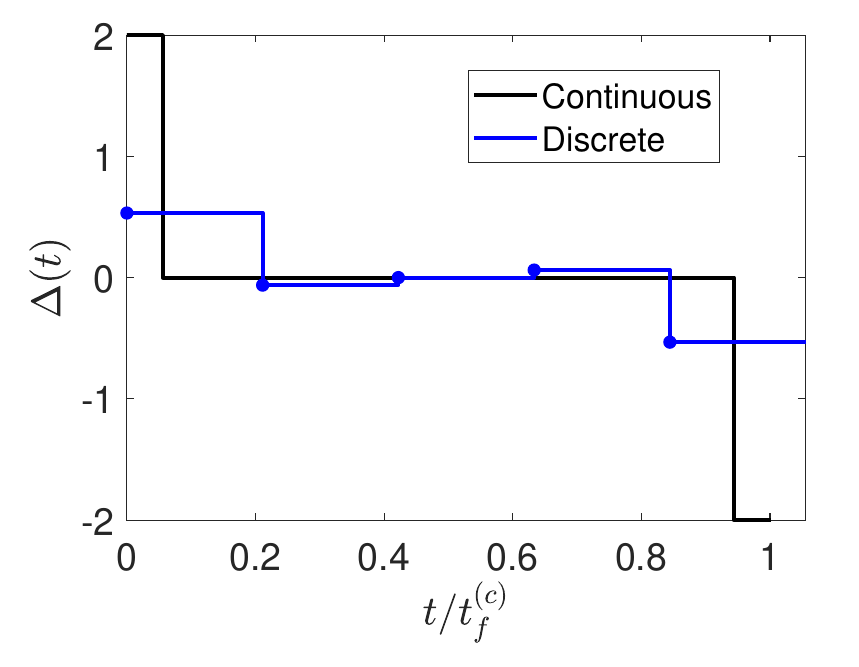}
  \caption{Plot of the optimal controls for the Landau-Zener Hamiltonian. The black and blue lines depict respectively the continuous and the discrete solutions for $N=5$. Quantities plotted are dimensionless.}
  \label{fig10}
\end{figure}

We analyze in this section a closely related control of a two-level quantum system that has been widely investigated both theoretically~\cite{hegerfeldt:2013,garcia2022,poggi2013} and experimentally~\cite{bason2012,zenesini2009,tayebirad2010} in recent years. The two levels are coupled through a Landau-Zener Hamiltonian leading to the dynamics of Eq.~\eqref{eqonecontrol} where $\omega$ is a constant parameter (set to 0.5 in the numerical computation) describing the coupling between the two levels and $\Delta(t)$ the control parameter. In the case of a Bose Einstein Condensate (BEC) in an optical lattice~\cite{bason2012}, this latter corresponds to the quasimomentum. The initial and final states of the control are respectively the ground adiabatic levels for $\Delta=+1$ and $\Delta=-1$. The two states are close to the north and south poles of the Bloch sphere. The time optimal solution in the continuous limit has been established in~\cite{hegerfeldt:2013}. With the constraint $|\Delta(t)|\leq \Delta_{\textrm{max}}$ (with $\Delta_{\textrm{max}}=2$ below), the control law is the concatenation of two bang pulses of maximum intensity at the start and end of the sequence with a zero singular control in the middle. It can be shown that this protocol reaches the QSL when the bang controls are replaced by Dirac pulses. Using the equations introduced in Sec.~\ref{sec4a}, where this time the role of control is played by $\Delta$, the same control problem can be solved with the constraint of piecewise constant pulses. This study is illustrated in Fig.~\ref{fig10} for the case $N=5$ and $\delta T=T$. We obtain  a ratio between the two control times equal to 1.0554. We surprisingly observe that the discrete solution does not use maximum intensity pulses for which $\Delta(t)=2$, but only less strong controls. As expected, the structure of the two optimal sequences are very close to each other.
\section{Time-optimal control in a linear system}\label{seclinear}
We have observed numerically in Sec.~\ref{sec3} and \ref{sec4} that the minimum control time by means of piecewise constant pulses converges exponentially towards the time in the continuous limit when the number of time steps $N$ tends to infinity. The aim of this section is to investigate the same issue but for a linearized quantum system. We show on a specific example that the linearization process is responsible for the change in convergence which is no more exponential, but polynomial with respect to the number of time steps.

A linearization of a quantum system can be achieved around an equilibrium point. This corresponds to the north or south pole of the Bloch sphere for a two-level quantum system. The control of quantum dynamics is then reduced to that of a spring which provides information on the initial problem~\cite{li2011,li2017,martikyan2020}. This approach is well-known in Nuclear Magnetic Resonance and usually called the \emph{small flip angle approximation}~\cite{bernstein2004}.

As a non-trivial example, we consider a linearization around the north pole of the quantum system investigated in Sec.~\ref{sec3}, but with non resonant controls. This slight modification is due to the fact that the optimal control is constant in the resonant case. In a given rotating frame, the dynamics of the two-level quantum system can be expressed in Bloch coordinates $(x,y,z)$ as
\begin{equation*}
\begin{split}
& \dot{x}=-\omega y+\sin(\varphi)z \\
& \dot{y}=\omega x-\cos(\varphi)z \\
& \dot{z}=-\sin(\varphi)x+\cos(\varphi)y
\end{split}
\end{equation*}
where $\omega$ is the offset term and $\varphi(t)$ the control parameter. Starting from the point $(0,0,1)$, the goal of the control is to reach a state in a neighborhood of the north pole. Since $z\simeq 1$ during the control process, we make this approximation in the dynamical system by replacing $z$ by 1 and by considering only the first two equations. The linearized system on $\mathbb{R}^2$ are then governed by the following differential equations
\begin{equation*}
\begin{split}
& \dot{x}=-\omega y+\sin\varphi \\
& \dot{y}=\omega x-\cos\varphi
\end{split}
\end{equation*}
In this case, the objective of the control is to steer in minimum time the system from the origin of coordinates $(0,0)$ to the point of coordinates $(1,0)$. We point out that the conclusion of this section would be the same with other initial and final states, as long as the linear approximation is valid. We first solve this problem in the continuous limit. The Pontryagin Hamiltonian $H_P$ is given by
$$
H_P=\omega(p_yx-p_xy)+p_x\sin\varphi-p_y\cos\varphi
$$
where the adjoint state $P=(p_x,p_y)$ is solution of $\dot{p}_x=-\omega p_y$ and $\dot{p}_y=\omega p_x$. We get
\begin{equation*}
\begin{split}
p_x(t)=p_x(0)\cos(\omega t)-p_y(0)\sin(\omega t) \\
p_y(t)=p_x(0)\sin(\omega t)+p_y(0)\cos(\omega t)
\end{split}
\end{equation*}
Note that $P^2=p_x^2+p_y^2$ is a constant of the motion. We introduce the angle $\vartheta$ such that $p_x=p\cos\vartheta$ and $p_y=p\sin\vartheta$, with $p$ the norm of $P$. The optimal control is given by $\frac{\partial H_P}{\partial \varphi}$ and fulfills $\tan\varphi = -\frac{p_x(t)}{p_y(t)}$. We arrive at $\tan\varphi=-\tan(\frac{\pi}{2}-\vartheta-\omega t)$ i.e.
$$
\varphi(t)=\vartheta+\omega t-\frac{\pi}{2}
$$
The optimal dynamics for the state of the system are then given by
$$
\dot{Z}=i\omega Z-e^{i\vartheta}e^{i\omega t}
$$
with $Z=x+iy$. The solution can be integrated exactly
$$
Z(t)=e^{i\omega t}[Z(0)-e^{i\vartheta}t]
$$
We stress that the optimal trajectories depend only on one free parameter $\vartheta$ which is fixed by the initial adjoint state. In the case $Z(0)=0$ and $Z_f=1$ under study, we deduce that
$$
e^{-i\omega t_f}=-t_fe^{i\vartheta}
$$
which leads to the minimum time $t_f^{(c)}=1$, and $\vartheta=\pi-\omega t_f$.

\begin{figure}[tb]
  \centering
  \includegraphics[width=1\linewidth]{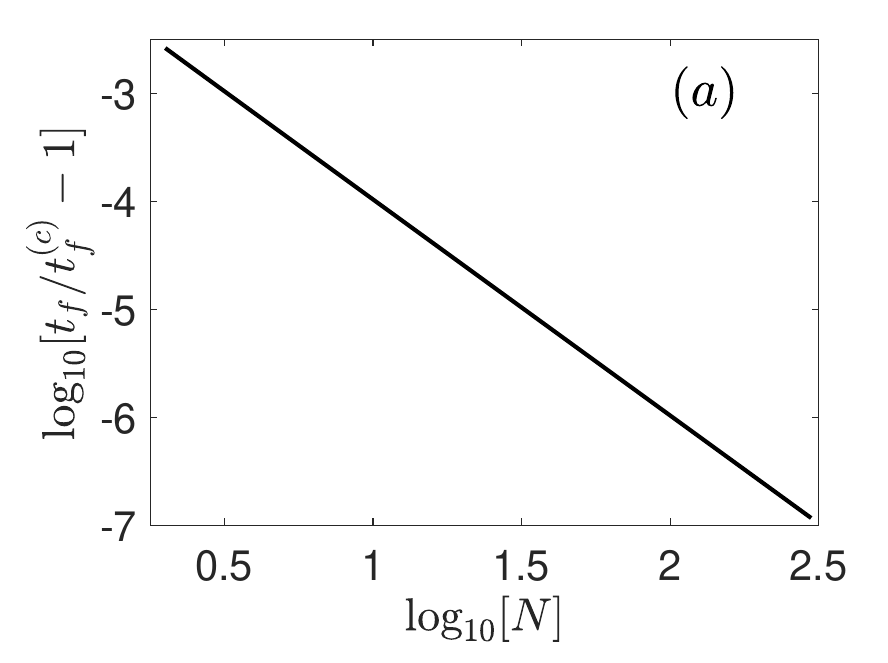}
  \includegraphics[width=1\linewidth]{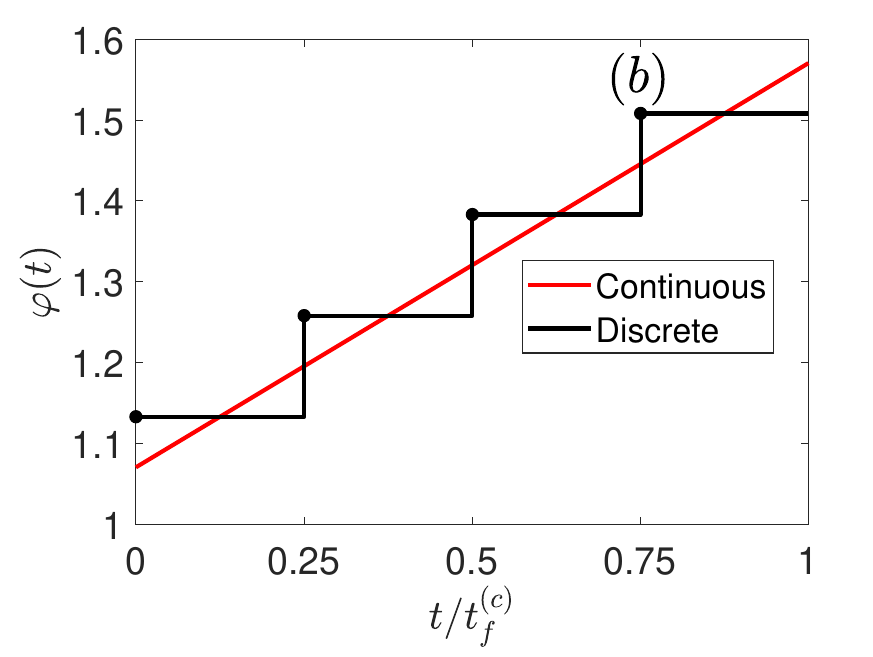}
   \caption{Evolution of the minimum time $t_f$ as a function of the number of steps $N$ (Panel (a)). The solid line is just to guide the eye. The panel (b) displays the optimal controls in the continuous limit (red or gray curve) and in the discrete case (black curve) for $N=4$. The parameter $\omega$ is set to 0.5 and the minimum time is equal to 1.0008 for $N=4$. Quantities plotted are dimensionless}
  \label{fig9}
\end{figure}

We consider now a piecewise constant control wit $t_f=NT$. The maximization condition is given by
$$
\tan\varphi_k=-\frac{\int_{kT}^{(k+1)T}p_x}{\int_{kT}^{(k+1)T}p_y}
$$
which leads after integration to
$$
\varphi_k=\vartheta+(k+\frac{1}{2})\omega T-\frac{\pi}{2}
$$
where $\vartheta$ has the same definition as in the continuous limit. This solution corresponds to the average of the continuous function $\varphi(t)$ in the interval $[kT,(k+1)T]$. The differential equation becomes in this case
$$
\dot{Z}=i\omega Z-e^{i\vartheta}e^{i(k+\frac{1}{2})\omega T}
$$
We have in the interval $[kT,(k+1)T]$
$$
Z_{k+1}=e^{i\omega T}Z_k-\frac{2}{\omega}\sin(\frac{\omega T}{2})e^{i(k+1)\omega T}e^{i\vartheta}
$$
with $Z_k=Z(kT)$. For the initial state $Z_0$, the final state is given by
$$
Z_N=e^{iN\omega T}Z_0-\frac{2N}{\omega}\sin(\frac{\omega T}{2})e^{i\vartheta}e^{iN\omega T}
$$
For the control problem, we deduce that
$$
e^{-iN\omega T}=-\frac{2N}{\omega}\sin(\frac{\omega T}{2})e^{i\vartheta}
$$
The parameter $T$ fulfills the condition
$$
T=\frac{2}{\omega}\arcsin(\frac{\omega}{2N})
$$
The total time is thus
$$
t_f=NT=\frac{2N}{\omega}\arcsin(\frac{\omega}{2N})
$$
The evolution of this time with respect to the minimum time in the continuous limit ($t_f^{(c)}=1$) when $N\gg 1$ is then
$$
\frac{t_f}{t_f^{(c)}}-1\simeq \frac{\omega^2}{24N^2}
$$
We show here that the convergence in this linear approximation is polynomial and not exponential as in the quantum case. These analytical results are illustrated in Fig.~\ref{fig9}. We conjecture that this speed of convergence is the same in any linear control system. The exponential convergence observed in quantum systems seems due to their bilinear structure, and therefore has an intrinsic link with the quantum nature of their dynamics.
\section{Gradient-based optimal control algorithm}\label{secgrape}
We describe in this section a formulation of gradient-based algorithm for piecewise constant pulses. We show how this algorithm is linked to the discrete version of the PMP.

\begin{figure}[tb]
  \centering
  \includegraphics[width=1\linewidth]{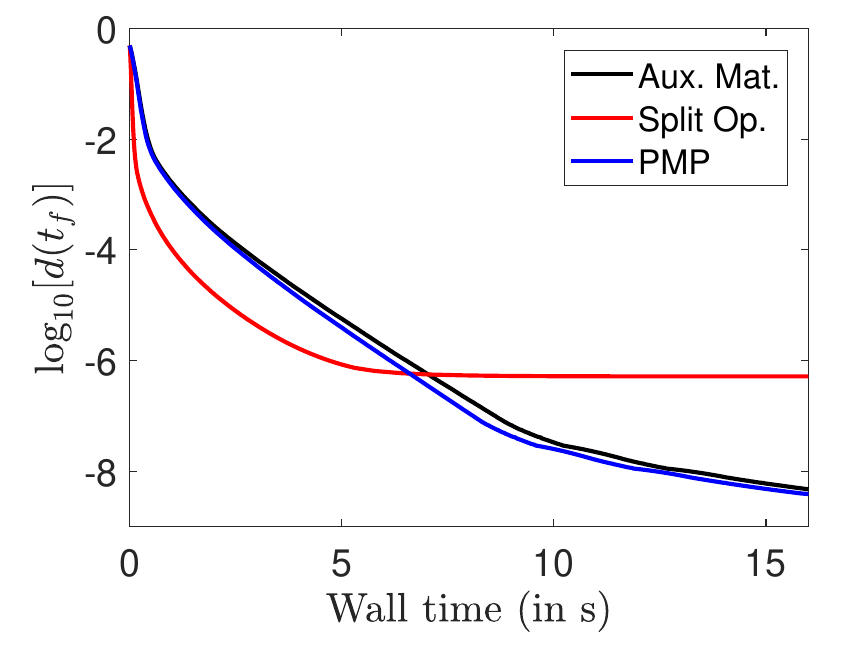}
  \caption{Comparison of the different formulations of GRAPE for the control problem described in Sec.~\ref{sec3}. The figure of merit $d$ is defined as $d(t_f)=1-|\langle\psi_f |\psi(t_f)\rangle|^2$. The red, black and blue lines depict the evolution of $d$ as a function of the wall time for respectively the split-operator, the auxiliary matrix and the discrete PMP approaches. The parameter $N$ is fixed to 7. Quantities plotted are dimensionless.}
  \label{fig8}
\end{figure}

We consider a general quantum system of state $|\psi(t)\rangle$ solution of the Schr\"odinger equation
$$
i|\dot{\psi}\rangle=[\cos(\varphi)H_x+\sin(\varphi)H_y]|\psi\rangle
$$
with $|\psi(0)\rangle=|\psi_0\rangle$ the initial state, the control being $\varphi(t)\in\mathbb{R}$. The optimal control problem is defined through the figure of merit $J$ to maximize
$$
J=|\langle \psi_f|\psi(t_f)\rangle |^2
$$
where $|\psi_f\rangle$ is the target state and the control time $t_f$ is fixed. The control parameter $\varphi$ is a piecewise constant function of values $(\varphi_k)$, $k\in \{0,1,\cdots, N-1\}$, with a sampling period $T$. The final state of the dynamics can be expressed as
$$
|\psi(t_f)\rangle=|\psi_N\rangle=U_{N-1}U_{N-2}\cdots U_0|\psi_0\rangle
$$
where $U_k=\exp[-i(\cos(\varphi_k)H_x+\sin(\varphi_k)H_y)T]$. We denote by $|\psi_k\rangle=U_{k-1}U_{k-2}\cdots U_0|\psi_0\rangle$ and $|\chi_k\rangle=U_{k}^\dagger U_{k+1}^\dagger\cdots U_{N-1}^\dagger|\psi_f\rangle$, the state and the adjoint state at step $k$. The figure of merit $J$ is then given by $J=|\langle \chi_{k+1}|U_k|\psi_{k}\rangle |^2$. The next step consists in computing the gradient of $J$ with respect to the control $\varphi$, which can be found from the derivative of $U_k$ with respect to $\varphi_k$. For that purpose, we use the Wilcox formula for the derivative of the exponential of a matrix~\cite{derivative}
$$
\frac{\partial e^{tA}}{\partial \theta}=e^{tA}\int_0^t e^{-t'A}\frac{\partial A}{\partial \theta}e^{t'A}dt'
$$
In our case, we get:
\begin{eqnarray*}
&  &\frac{\partial}{\partial \varphi_k}\exp[-i\tau(\cos(\varphi_k)H_x+\sin(\varphi_k)H_y)]=\\
& & U_k(\tau)\int_0^\tau U_k^\dagger(t)(-i)[-\sin(\varphi_k)H_x+\cos(\varphi_k)H_y]U_k (t)dt
\end{eqnarray*}
which leads to
\begin{eqnarray*}
& & \frac{\partial J}{\partial \varphi_k}=\\
& & -2\sin(\varphi_k)\Im[\langle\psi_N|\psi_f\rangle\langle\chi_{k}|\int_0^\tau U_k^\dagger(t)H_xU_k (t)dt|\psi_{k}\rangle]\\
& & +2\cos(\varphi_k)\Im[\langle\psi_N|\psi_f\rangle\langle\chi_{k}|\int_0^\tau U_k^\dagger(t)H_y U_k (t)dt|\psi_{k}\rangle]
\end{eqnarray*}
and finally to
\begin{eqnarray*}
& & \frac{\partial J}{\partial \varphi_k}=2\int_0^\tau \Im[\langle\psi_N|\psi_f\rangle\langle\chi_{k}(t)[-\sin(\varphi_k)H_x\\
& & +\cos(\varphi_k)H_y]|\psi_{k}(t)\rangle ] dt
\end{eqnarray*}
or in a compact form:
$$
\frac{\partial J}{\partial \varphi_k}=-\sin(\varphi_k)I_x^{(k)}+\cos(\varphi_k)I_y^{(k)}
$$
with
\begin{equation*}
\begin{split}
& I_x^{(k)}=2\int_0^\tau \Im[\langle\psi_N|\psi_f\rangle\langle\chi_{k}(t)|H_x|\psi_{k}(t)\rangle ] dt \\
& I_y^{(k)}=2\int_0^\tau \Im[\langle\psi_N|\psi_f\rangle\langle\chi_{k}(t)|H_y|\psi_{k}(t)\rangle ] dt
\end{split}
\end{equation*}
The maximization condition is achieved when all the gradients are zero. We recover here a formula for the optimal control very similar as the one given by the discrete version of the PMP
\begin{equation}\label{eqgrad}
\tan(\varphi_k)=\frac{I_y^{(k)}}{I_x^{(k)}}
\end{equation}
This derivation gives a qualitative justification of the extended PMP formulation for piecewise constant pulses. Note that this argument can be applied to other expressions of the Hamiltonian of the system. Equation~\eqref{eqgrad} also allows to define a new version of GRAPE which has the key advantage of giving an exact value of the gradient~\cite{boscain21}. We now study the control problem described in Sec.~\ref{sec3} with $|\psi(0)\rangle=\frac{1}{\sqrt{2}}[|1\rangle+|2\rangle]$ and $|\psi_f\rangle=\frac{1}{\sqrt{2}}[|1\rangle+i|2\rangle]$ where $\{|1\rangle,|2\rangle\}$ is the canonical basis of the Hilbert space. We compare in Fig.~\ref{fig8} this formulation to the standard one based on a split-operator method~\cite{grape} and to the recent auxiliary matrix approach~\cite{goodwin2015}. We point out that this latter method gives also an exact gradient but at the cost of increasing by a factor 2 the size of the system, while the first only approximates the gradient of $J$. In order to get a fair comparison between the algorithms, the same parameters in the optimization procedure have been used. We have checked that the qualitative conclusions do not depend on a specific choice of the numerical parameters. We choose to plot the fidelity $d(t_f)=1-|\langle\psi_f |\psi(t_f)\rangle|^2$ as a function of the wall time. Similar results
have been achieved for the CPU time. Note that the algorithms based on the PMP and on the auxiliary matrix method give exactly the same performance when the figure of merit is plotted in terms of the number of iterations. This observation was expected since the gradient is computed without any approximation in the two cases. The slight difference observed with the wall time is due to the computation time for calculating the gradient. Here the PMP version gives better results because the integrals involved in Eq.~\eqref{eqgrad} can be determined exactly. However, the auxiliary matrix approach seems to be more efficient in terms of computational time if applied to large quantum systems. This aspect which is beyond the scope of this paper should be an interesting point to check. A comparison with an automatic differentiation method~\cite{AD1,AD2,AD3} has also been performed. The same final distance to the target state is reached but with longer computation times than the PMP or the auxilary matrix method. Another key advantage of the PMP formulation of GRAPE lies in the fact that it is not restricted to bilinear dynamics, as it is the case, e.g., for the auxiliary matrix approach. The same algorithm can be used if nonlinearities are added to the model system~\cite{lapert08,ohtsuki08,jager2014,borzi2007,hocker2016}.

\begin{figure}[tb]
  \centering
  \includegraphics[width=1\linewidth]{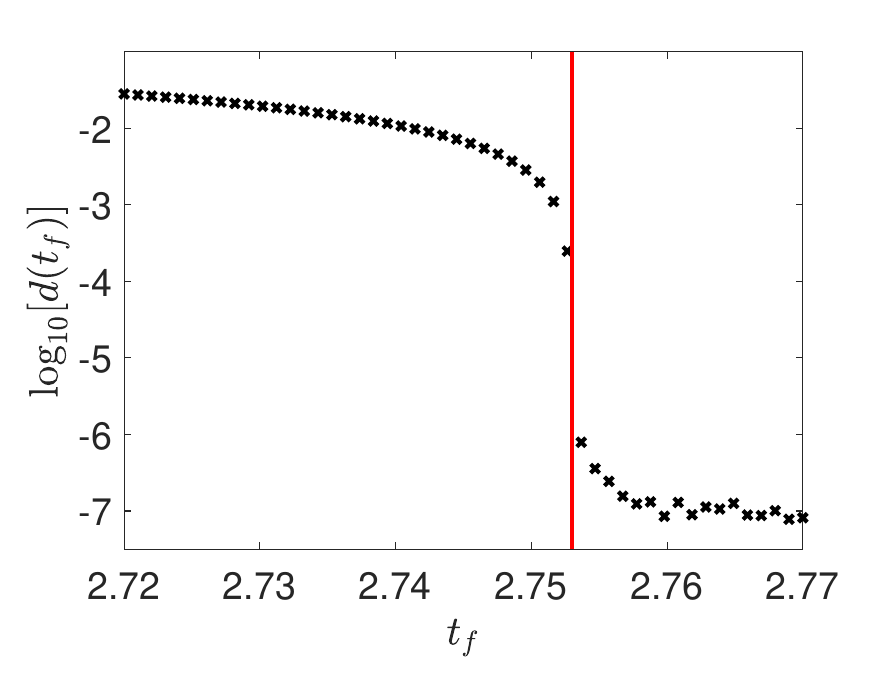}
  \caption{Numerical results (crosses) obtained with GRAPE for $N=3$ and $\delta T=T$ in the case of the process with two controls. The control times range from 2.72 to 2.77. The vertical solid red line indicates the value of the minimum time found by this optimization procedure. Quantities plotted are dimensionless.}
  \label{fig5}
\end{figure}

Finally, we propose a numerical analysis of the minimum time with the GRAPE algorithm based on the auxiliary matrix approach. Note that the same results in terms of precision are obtained with the PMP formulation of GRAPE. For a fixed control time $t_f$, the goal of the optimization procedure is to minimize the figure of merit $d(t_f)$. A good estimation of the optimal solution can be achieved
by running the algorithm with many different initializations. This operation is repeated for a range of final times $t_f$ in order to estimate the minimum time which is the lower value of $t_f$ for which $d(t_f)\simeq 0$. In the numerical simulations displayed in Fig.~\ref{fig5}, we observe that the figure of
merit $d$ falls very quickly towards zero around $t_f = 2.753$, which is the minimum time of the control process. The distance to the target is of the order of $10^{-6}$. For the shooting algorithm, we obtain in the same conditions a minimum time of 2.75292 and a distance of $10^{-9}$.


\section{Conclusion}\label{sec5}
We have applied an extension of the PMP to piecewise constant pulses to the control in minimum time of two-level quantum systems. We consider as illustrative examples two benchmark control problems for which the control procedure is known exactly in the continuous limit. We compare for different state-to-state transfers the minimum times, the optimal trajectories and the controls. We show in particular that the minimum time in the discrete case converges very quickly towards its continuous limit, while the piecewise and the continuous controls remain quite different. In the examples under study, we observe that this convergence is exponential. These results can be considered as good news from an experimental point of view since experimental sampling limitations only slightly affect the minimum control time and therefore the overall duration of the operation composed of several control processes.

Finally, these results pave also the
way for other studies using the same approach, such as the
generation of one-qubit gates~\cite{kozbar:2012,garon,boozer:2012}, the control of open quantum systems~\cite{khaneja2003,mukherjee2013,lapert:2013,lapertprl}, the selective~\cite{silver:1984,silver:1985,vandamme:2018,basilewitsch:2020} and robust~\cite{vandamme:2017,daemsprl,STAnjp,compvitanov,zeng2018,zeng2019} control of two-level quantum systems or the control of more complex quantum systems~\cite{roadmap,dupont21,lapert2012b,nimbalkar:2012}. It would be interesting to verify in these examples that the convergence is always exponential and does not depend on the control problem under study.

\noindent\textbf{Acknowledgment}\\
This research has been supported by the ANR project ``QuCoBEC'' ANR-22-CE47-0008-02. We gratefully acknowledge useful discussions with J. Billy, D. Gu\'ery-Odelin and B. Peaudecerf, who suggested the idea of this paper.\\

\appendix

\section{The shooting method in the case of two controls}\label{appA}
\begin{figure}[tb]
  \centering
  \includegraphics[width=1\linewidth]{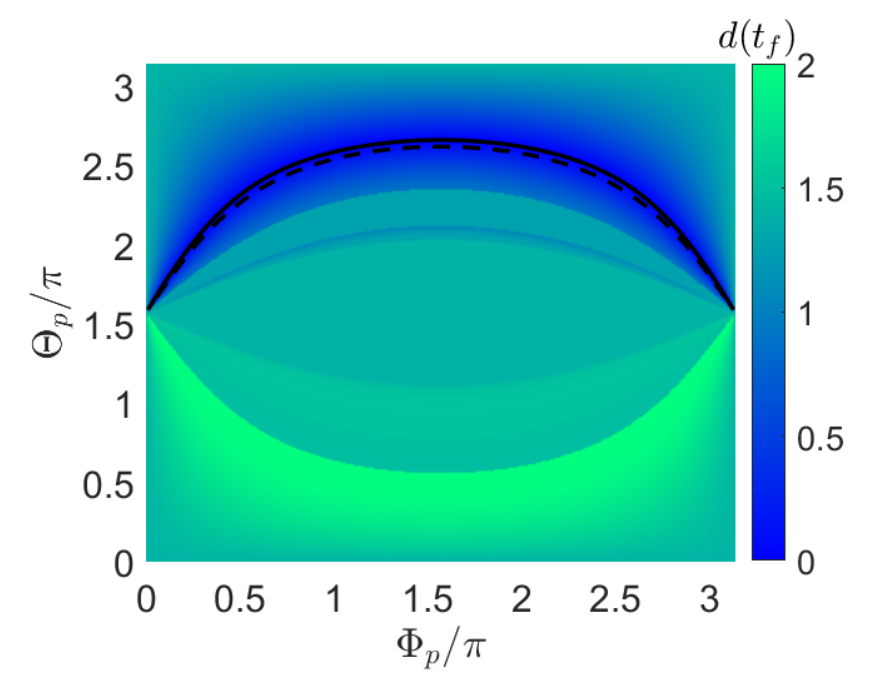}
  \caption{Figure of merit $d$ for the minimum discrete time $t_f$ ($N=3$ and $\delta T=T$) as a function of the coordinates $(\Theta_p,\Phi_p)$ of the initial adjoint state (see the text for details). The solid and dashed lines depict respectively the minimum value of $d$ and the continuous limit. Quantities plotted are dimensionless.}
  \label{fig6}
\end{figure}

We describe in this section the solutions of the shooting algorithm in the case of two controls. We first recall some results about the continuous limit~\cite{boscain21,sugny:2008}. In this limit, the optimal equations can be integrated exactly by introducing spherical coordinates $(r,\vartheta,\varphi)$ such that
$x=r\sin\vartheta\cos\varphi$, $y=r\sin\vartheta\sin\varphi$ and $z=r\cos\vartheta$, with by definition $r=1$. Using the generating function~\cite{goldstein} $F_2=p_xr\sin\vartheta\cos\varphi+p_yr\sin\vartheta\sin\varphi+p_zr\cos\vartheta$, we deduce that the conjugate momenta can be expressed as
\begin{eqnarray*}
& & p_r=p_x\sin\vartheta\cos\varphi+p_y\sin\vartheta\sin\varphi+p_z\cos\vartheta\\
& & p_\vartheta=p_x\cos\vartheta\cos\varphi+p_yr\cos\vartheta\sin\varphi-p_zr\sin\vartheta \\
& & p_\varphi=-p_xr\sin\vartheta\sin\varphi+p_yr\sin\vartheta\cos\varphi
\end{eqnarray*}
It is straightforward to show that $p_x(0)=p_r(0)$, $p_y(0)=p_\varphi(0)$ and $p_z(0)=-p_\theta(0)$ since $\vartheta(0)=\frac{\pi}{2}$ and $\varphi(0)=0$. In the spherical coordinates, the true Pontryagin Hamiltonian (obtained by replacing the controls by their values maximizing $H_P$) can be written as
$$
H_P=\sqrt{p_\vartheta^2+p_\varphi^2/\tan^2\vartheta}
$$
with the constraint $H_P=1$ at any time. The Hamiltonian's equations then read
\begin{eqnarray*}
& & \dot{\vartheta}=p_\vartheta;~\dot{\varphi}=p_\varphi/\tan^2\vartheta \\
& & \dot{p}_\vartheta=p^2_\varphi/\tan\vartheta(1+\frac{1}{\tan^2\vartheta});~\dot{p}_\varphi=0 \\
\end{eqnarray*}
At time $t=0$, using $\vartheta(0)=\frac{\pi}{2}$ and $H_P=1$, we deduce that $p_\vartheta (0)=\pm 1$. The extrema of the trajectory are reached when $t=\frac{\pi\sqrt{3}}{4}$ and $\dot{\vartheta}=0$. Using $\dot{\vartheta}^2=1-\frac{p_\varphi^2}{\tan^2\vartheta}$, we deduce that the corresponding angle $\vartheta_m$ satisfies $\tan\vartheta_m=p_\varphi$. Starting from the time evolution of $\vartheta$ given, when $\vartheta(t)$ increases, by~\cite{sugny:2008}
$$
\vartheta(t)=\frac{\pi}{2}+\textrm{asin}[\frac{1}{\sqrt{1+p_\varphi^2}}\sin(\sqrt{1+p_\varphi^2}t),
$$
we arrive at $p_\varphi=\frac{1}{\sqrt{3}}$. Note that the value of $p_r$ does not play any role in the control process. Finally, we obtain the possible values of the initial adjoint state $P(0)=(p_x(0),\frac{1}{\sqrt{3}},\pm 1)$ given in the main text of the paper. We choose to use a sphere to represent the different initial adjoint states with the following spherical coordinates $p_x=R_p\sin\Theta_p\cos\Phi_p$, $p_y=R_p\sin\Theta_p\sin\Phi_p$ and $p_z=R_p\cos\Theta_p$. The constant radius is equal to $R_p=\sqrt{\frac{4}{3}+p_x^2(0)}$ and the angles $\Theta_p$ and $\Phi_p$ can be expressed when $p_z(0)=-1$ as
\begin{eqnarray*}
& &\Theta_p=\textrm{acos} [-1/\sqrt{\frac{4}{3}+p_x^2(0)}] \\
& & \Phi_p=\textrm{atan} [\frac{1}{\sqrt{3}p_x(0)}]
\end{eqnarray*}
The corresponding curve is plotted in Fig.~\ref{fig6}. The result of the control process for the discrete case can also be represented on the same sphere. To this aim, we consider all the possible values of $P(0)$ and we integrate the trajectory up to the minimum time $t_f$. The figure of merit $d$ is defined as the Euclidian distance from the final state to the target. A contour plot of the corresponding performance is displayed in Fig.~\ref{fig6}. The optimal adjoint states correspond to $d=0$. We observe that they also form a curve in the space $(\Theta_p,\Phi_p)$ which is very close to the one of the continuous case. Note that all these initial adjoint states lead to the same controls in the discrete version of the PMP.

\section{Application of the PMP in the case of one control}\label{appB}
We show in this section how the PMP in the discrete case can be applied to the control problem of Sec.~\ref{sec4}. We introduce the switching function $\Phi=p_zy-p_yz$. We denote by $\Phi_k(t)$ the value of the switching function in the interval $[kT,(k+1)T]$. The maximization condition can be written as follows:
$$
(\mathcal{U}-\omega_k)\frac{1}{T}\int_{kT}^{(k+1)T}\Phi_k(t)dt\leq 0,~\forall \mathcal{U}\in [-1,1]
$$
where $\omega_k$ is the constant value of $\omega(t)$ in this interval. We have the following cases, with $\Gamma_k(\omega_k)=\frac{1}{T}\int_{kT}^{(k+1)T}\Phi_k(t)dt$:
\begin{itemize}
\item If $\min_{\omega_k}\Gamma_k>0$ then $\Gamma_k(\omega_k)>0$ and $\mathcal{U}-\omega_k\leq 0$. We deduce that $\omega_k=+1$.
\item If $\max_{\omega_k}\Gamma_k<0$ then $\Gamma_k(\omega_k)<0$ and $\mathcal{U}-\omega_k\geq 0$. We deduce that $\omega_k=-1$.
\item If $\max_{\omega_k}\Gamma_k>0$ and $\min_{\omega_k}\Gamma_k<0$ then the only solution is $\Gamma_k(\omega_k)=0$ which leads to a value $\omega_k$ in the interval $[-1,1]$.
\end{itemize}
Note that it is possible to get solutions distinct from -1 or +1 in the discrete case while the optimal trajectory is regular in the continuous limit, i.e. $|\omega(t)|=1$ except in isolated times.

The expression of $\Gamma$ can be derived explicitly as follows. To simplify the notations, we consider a time interval $[0,T]$ such as the state and the adjoint state are respectively of coordinates $(x_0,y_0,z_0)$ and $(p_{x0},p_{y0},p_{z0})$ at $t=0$. We also introduce the coordinates of the angular momentum $L=X\times P$
\begin{eqnarray*}
& & L_x=yp_z-zp_y=\Phi \\
& & L_y=zp_x-xp_z \\
& & L_z=xp_y-yp_x
\end{eqnarray*}
We have
\begin{eqnarray*}
& & x(t)=x_0-\frac{\Delta B}{\Omega_0}-\frac{\Delta}{\Omega_0}(A\sin(\Omega_0 t)-B\cos(\Omega_0 t)) \\
& & y(t)=A\cos(\Omega t)+B\sin(\Omega t) \\
& & z(t)=z_0+\frac{\omega B}{\Omega_0}+\frac{\omega}{\Omega_0}(A\sin(\Omega_0 t)-B\cos(\Omega_0 t))
\end{eqnarray*}
with $A=y_0$, $B=\frac{\Delta x_0-\omega z_0}{\Omega_0}$ and $\Omega_0=\sqrt{\omega^2+\Delta^2}$. For the adjoint state, we get similar equations
\begin{eqnarray*}
& & p_x(t)=p_{x0}-\frac{\Delta D}{\Omega_0}-\frac{\Delta}{\Omega_0}(C\sin(\Omega_0 t)-D\cos(\Omega_0 t)) \\
& & p_y(t)=C\cos(\Omega_0 t)+D\sin(\Omega_0 t) \\
& & p_z(t)=p_{z0}+\frac{\omega D}{\Omega_0}+\frac{\omega}{\Omega_0}(C\sin(\Omega_0 t)-D\cos(\Omega_0 t))
\end{eqnarray*}
with $C=p_{y0}$ and $D=\frac{\Delta p_{x0}-up_{z0}}{\Omega_0}$.

The switching function can then be expressed as
\begin{eqnarray*}
& & \Phi=\frac{\omega}{\Omega_0^2}(\Delta L_{z0}+\omega L_{x0})+\frac{\Delta}{\Omega_0^2}(\Delta L_{x0}-\omega L_{z0})\cos(\Omega_0 t) \\
& & -\frac{\Delta}{\Omega_0}L_{y0}\sin(\Omega_0 t)
\end{eqnarray*}
The corresponding function $\Gamma=\frac{1}{T}\int_0^T\Phi(t)dt$ is given by
\begin{eqnarray*}
& & \Gamma=\frac{\omega}{\Omega_0^2}(\Delta L_{z0}+\omega L_{x0})+\frac{\Delta}{\Omega_0^2} (\Delta L_{x0}-\omega L_{z0})\textrm{sinc}(\Omega_0 T)\\
& & +\frac{\Delta}{\Omega_0}L_{y0}\frac{\cos(\Omega_0 T)-1}{\Omega_0 T}
\end{eqnarray*}
As expected, we obtain in the limit $T\to 0$ that $\Gamma\simeq \Phi$. In the general case, $\Gamma$ is a function of $\omega$ in the interval $[-1,1]$. At each time step, we compute numerically the maximum, the minimum and the zeros of $\Gamma$ to find the corresponding value of $\omega$.

\end{document}